\documentclass[9pt,twocolumn,twoside]{osajnl}

\journal{ol} 

\setboolean{shortarticle}{true}
\usepackage{graphicx} 
\usepackage{amsmath,amssymb}
\usepackage{caption}
\usepackage{subcaption}

\title{Polarization-Based Truncated SU(1,1) Interferometer based on Four-wave Mixing in Rb vapor}

\author[1*]{Nikunjkumar Prajapati}
\author[1]{Irina Novikova}

\affil[1]{Physics Department, William $\&$ Mary, 300 Ukrop Way, Willaimsbrg VA, 23187}

\affil[*]{Corresponding author: nprajapati@email.wm.edu}

\begin{abstract}
We propose and demonstrate a polarization-based truncated SU(1,1) interferometer that outputs the desired optical joint-quadrature of a two-mode squeezed vacuum field and allows its measurements using a single balanced homodyne detector. Using such setup we demonstrated up to $\approx 2$~dB of quantum noise suppression below the shot-noise limit in intensity-difference and phase-sum joint quadratures, and confirmed entanglement between the two quantum fields. Our proposed technique results in a better balance between the two ports of the detector and, consequently, in better common noise suppression for differential measurements. As a result, we were able to observe flat joint-quadrature squeezing and entanglement at wide range of detection frequencies: from several MHz (limited by the photodiode gain bandwidth) down to a few hundred Hz (limited by electronic noises).
\end{abstract}

\setboolean{displaycopyright}{true}
\begin{document}
	
\maketitle

Precision interferometric phase measurements are the basis of many optical instruments, from gravitational wave detection~\cite{GW_detect_with_squeezed:Chau:2014,quant_sense:pooser:2018} to biosensing~\cite{bio_inter:Qin:16,quant_sense:pooser:2018} and quantum information~\cite{bell_states,quant_sense:pooser:2018,quant_metrol:Zhang:2014,quant_inter:Kimble:2008}. The sensitivity of classical interferometers is ultimately limited by the optical shot noise~\cite{int_reviewPRA1986,Chekhova:16AOPreview,Andersen_2016,PhysRevA.82.033819}. However, this limitation can be overcome in a SU(1,1) interferometer in which classical beam splitters are replaced by non-linear beam splitters (NLBS), where the input/output ports are correlated via strong nonlinearity~\cite{int_reviewPRA1986,Chekhova:16AOPreview,zhangAPL2011_SU11,CheckovaPhysRevLett.119.223604,Zhang18OL}. In this arrangement two-mode squeezed vacuum optical fields are produced at the first NLBS and then interfered at the second NLBS, in principle achieving Hisenberg-limited relative phase sensitivity. Similar performance is possible in a truncated SU(1,1) interferometer, in which the second NLBS is replaced by full homodyne detection of each optical field~\cite{PhysRevA.95.063843Lett,Anderson:17Optica,quant_sense:pooser:2018}. 

\begin{figure*}[!bhpt]
	\centering
	\includegraphics[width=1\textwidth]{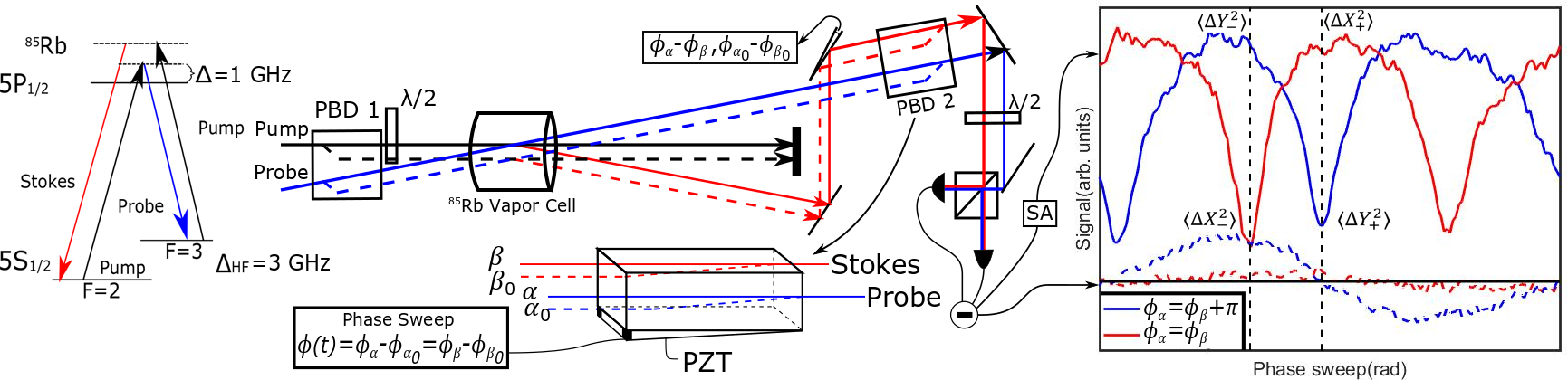}
	\caption{\emph{(a)} Level diagram of the  FWM process in which the pump (black), probe (blue), and Stokes (red) optical fields are in a four-photon resonance with the two hyperfine sublevels of the $5S_{1/2}$ state of ${}^{85}$Rb connected through a virtual state detuned by $\approx +1$~GHz from the 5$P_{1/2}$ state (pump beam at 794.9727~nm). (b) The experimental geometry of the dual-rail polarization-based truncated SU(1,1) interferometer. (c) Example of the experimental noise power (solid) and differential intensity (dashed) signals plotted as a function of the relative phase between each of the local oscillators (LOs) and the corresponding squeezed fields $\phi(t)$ varying between $0$ and  $2\pi$. The blue curves show the case where $\phi_{\alpha}=\phi_{\beta}+\pi$ and the red curves show the case where $\phi_{\alpha}=\phi_{\beta}$.}
	\label{fig:SU11}
\end{figure*}

Here we present a polarization-based realization of the truncated SU(1,1) interferometer that outputs the selected joint quadratures of the two-mode interferometer by combining the orthogonally-polarized signal quantum fields and local oscillators. For the realization of the nonlinear beam splitter in our experiment, we utilize four-wave mixing (FWM) in hot ${}^{85}$Rb vapor~\cite{lettPRA08,lettSci08,zhangAPL2011_SU11,Prajapati19OL}: a nonlinear process which allows for strong phase-sensitive amplification of the probe and the generation of the Stokes optical fields which makes them quantum-correlated. In our experiment, we use  a two-rail geometry to generate the quantum correlated two-mode squeezed fields and the local oscillators (LOs) in the same vapor cell via the FWM process. The two rails have orthogonal polarizations which makes it easy to recombine the beams from the two rails and control their relative phase.

The proposed arrangement has many advantages. First, it enables direct control of the detected joint-quadratures by adjusting the relative phase between the probe and Stokes optical fields and their respective local oscillators (LOs) rather than electronically adding or subtracting the individually-measured signals. Second, the improved symmetry and stability of our configuration removes the need for post measurement balancing as done in ~\cite{LettOE2018,quant_sense:pooser:2018} and allows efficient broad-band operation with the detection frequencies from as low as 200 Hz to above a few MHz. Finally, the same setup can be used for the generation of polarization-entangled Bell states~\cite{bell_states,Park2019PolarizationEntangledPF}.

The realization of the polarization-based truncated SU(1,1) interferometer using the four-wave mixing in a ${}^{85}$Rb vapor cell is shown in Fig~\ref{fig:SU11}. Both pump and probe input fields are derived from the same amplified diode laser system (see Ref.\cite{Prajapati19OL} for details) and then combined at a 4~mrad angle using an edge mirror.
After that, a single Wollaston polarization beam displacer (PBD1) was used to split each of the two beams into the two rails, separated vertically by approximately 4~mm. The pump beam was split evenly between the two rails and had a power of 160~mW for each channel while the probe had different powers for each rail. For the LO channel a probe field ($\approx 30~\mu$W) was injected into the upper rail, producing approximately $1$~mW combined power for the probe and Stokes fields after the cell. The bottom rail contained the analyzed two-mode squeezed vacuum fields and thus was seeded with either vacuum or very weak ($<1~\mu$W) coherent field for all squeezing measurements. After the displacer the two pump beam polarizations were rotated by $90^\circ$ with respect to those of the probes using an edge-mounted half wave plate before entering a 25~mm long ${}^{85}$Rb vapor cell enclosed in a three layer magnetic sheilding. The cell was maintained at $106^\circ C$. At the location of the cell, the probe beam waist were ~300~$\mu$m and the pump beam waist was ~500~$\mu$m. 

After the cell, the pump optical fields were blocked using an opaque mask, while the quantum signals and local oscillator beams were combined using the second beam displacer (PBD2). Finally, the resulting probe and Stokes beams polarizations were rotated by $\pi/2$ using a $\lambda/2$ wave plate, so they could be evenly mixed using a regular polarization beam-splitter (PBS) before detection at the balanced photodetector (BPhD), thus realizing a standard homodyne detection scheme. To ensure good spatial mode overlap between the probe and Stokes fields in two rails, we verified that the visibility of the interference fringes in case of equal input probe fields was better than $98~\%$.

To illustrate the measurement principle mathematically, we calculated the instantaneous photocurrents at each BPhD, by labeling the probe fields in the upper and bottom rails as $\hat\alpha$ and $\hat\alpha_0$, and the Stokes fields -- $\hat\beta$ and $\hat\beta_0$ correspondingly:
\begin{eqnarray}
\hat i_1 &\propto &  \left|\hat\alpha+\hat\alpha_0+\hat\beta-\hat\beta_0\right|^2;\label{eq:BPDfields1} \\
\hat i_2 &\propto &  \left|\hat\beta+\hat\beta_0-\hat\alpha+\hat\alpha_0\right|^2. \nonumber
\label{eq:BPDfields}
\end{eqnarray}
Here each field can be presented in the form $\hat\alpha=(\alpha+\delta \hat\hat X_\alpha+i\delta \hat Y_\alpha)e^{i\phi_\alpha}$~\cite{scullybook}, where $\alpha=\langle \hat\alpha\rangle$ is the mean amplitude, $\delta \hat\hat X_\alpha$ and $\delta \hat Y_\alpha$ are the quantum noise quadratures, and $\phi_\alpha$ is the phase of the corresponding optical field. 

Subtracting the two photo currents in Eqs.(\ref{eq:BPDfields1})  and by assuming the amplitudes for the local oscillators  are similar and much larger than the amplitudes of the squeezed beams ($\alpha\simeq \beta \gg \alpha_0, \beta_0$), we can calculate the instantaneous differential photo current: 

\begin{equation}\label{eq:Delta_i}
\begin{split}
\hat i_- & =\alpha \alpha_0 \cos(\phi(t)-\phi_\alpha)-\beta \beta_0 \cos(\phi(t)-\phi_\beta)\\
&+\alpha(\delta \hat X_{\alpha_0}\cos(\phi(t)-\phi_\alpha)-\delta \hat Y_{\alpha_0}\sin(\phi(t)-\phi_\alpha))\\
&-\beta(\delta \hat X_{\beta_0}\cos(\phi(t)-\phi_\beta)+\delta \hat Y_{\beta_0}\sin(\phi(t)-\phi_\beta)),
\end{split}
\end{equation}
where $\phi(t) =  \phi_{\alpha_0}-\phi_{\alpha} = \phi_{\beta_0}- \phi_{\beta}$ is the  time-varying phase between the probe (or Stokes) fields and the corresponding local oscillators.  
The photo current variance $\langle (\Delta i)^2\rangle$ is proportional to the noise power measured by the spectrum analyzer in a standard homodyne detection scheme, and thus was used to characterize the joint noise quadratures of the two-mode squeezed vacuum fields $\hat\alpha_0$ and $\hat\beta_0$.

The mean value of the differential current $\langle i_- \rangle$ describes the classical intensity difference between the two ports of the balanced detector. Depending on the input of the quantum enhanced channel for the probe, either weak seed (approximately $1~\mu W$) or vacuum probe beam, two different signals will be observed. 
In the presence of a weak coherent probe field, then $\langle i_- \rangle$ represents the output of a classical interferometer while still maintaining the quantum enhanced noise properties. In the absence of a probe beam, the signal ($\langle i_- \rangle$) is a flat line while the noise spectrum still moves through the joint quadratures as the phases are changed. In this experiment, a seed field of $<1~\mu W$ was always maintained to be able to track the various phases.

The key advantage of this configuration is its capability to optically control the relative phases between the two-mode squeezed fields and the local oscillator such that  both phase and amplitude joint quadratures $\hat X_\pm$ and $\hat Y_\pm$ can be measured:
\begin{eqnarray}
\hat X_\pm &=&   \delta \hat X_{\alpha_0}\pm\delta \hat X_{\beta_0}  ,\\
\hat Y_\pm &=&   \delta \hat Y_{\alpha_0}\pm\delta \hat Y_{\beta_0}  .
\end{eqnarray}

For example, we can simultaneously change the relative phase $\phi(t)$ of both local oscillators  with respect to their corresponding squeezed fields by fine-tuning the vertical alignment of the PBD. This is a result of the refractive index differences of the two orthogonal polarizations. In the measured experimental spectra we were able to continuously sweep the value of this phase using the BPD2 mounted in a piezo-electric transducer (PZT) optical mount. Importantly, such adjustments do not result in any noticeable deterioration of the interference fringe visibility between the two inputs. 

The additional joint quadrature selection was achieved by varying the relative phase between the two local oscillators, $\phi_\alpha$ and $\phi_\beta$, using small mirror displacements in the Stokes channel. Let us first analyze the case of the two LOs being in phase with one another ($\phi_\alpha=\phi_\beta$). In this case, the classical differential intensity ($\langle i_- \rangle = (\alpha \alpha_0 -\beta \beta_0) \cos(\phi(t)-\phi_\alpha)$) is zero for all values of $\phi(t)$ for balanced probe/Stokes fields ($\alpha\approx\beta$ and $\alpha_0 \approx \beta_0$). This is because for each of the photodetectors, the interference maxima of the two probe fields occur at the same phase $\phi(t)$ as the interference minima of the Stokes fields (and vice versa), and thus the total intensity at each photodetector is constant. At the same time, the measured joint-quadrature noise power is defined as $ \langle {\Delta \hat i}^2\rangle = \langle {i_-}^2 \rangle - {\langle i_- \rangle}^2$:
\begin{equation}\label{eq:phase_zero}
\begin{split}
\langle\Delta \hat i^2\rangle =&\alpha^2[\cos^2(\phi(t))\langle(\delta \hat X_{\alpha_0}-\delta \hat X_{\beta_0})^2\rangle\\
&+\sin^2(\phi(t))\langle(\delta \hat Y_{\alpha_0}-\delta \hat Y_{\beta_0})^2\rangle]
\end{split}
\end{equation}
Because of the nature of the FWM process, we expect to observe the two-mode squeezing in the intensity-difference ($\langle \Delta \hat X_-^2 \rangle$) and phase-sum noise quadratures ($\langle \Delta \hat Y_+^2 \rangle$), and two-mode anti-squeezing in the intensity-sum ($\langle \Delta \hat X_+^2 \rangle$) and phase-difference noise quadratures ($\langle \Delta \hat Y_-^2 \rangle$). Thus, for this experimental configuration, the sweep of the local oscillators relative phase $\phi(t)$ should oscillate between the lowest (squeezed) values for the intensity-difference quadrature ($\langle \Delta \hat X_-^2 \rangle$) and the highest (anti-squeezed) values for the phase-difference ($\langle \Delta \hat Y_-^2 \rangle$) quadrature. This behavior is shown experimentally in Fig.~\ref{fig:SU11}(c); by the red dashed and solid lines for the differential intensity and joint-quadrature noise sweeps, respectively.

\begin{figure}[htbp]
	\centering
	\includegraphics[width=1\columnwidth]{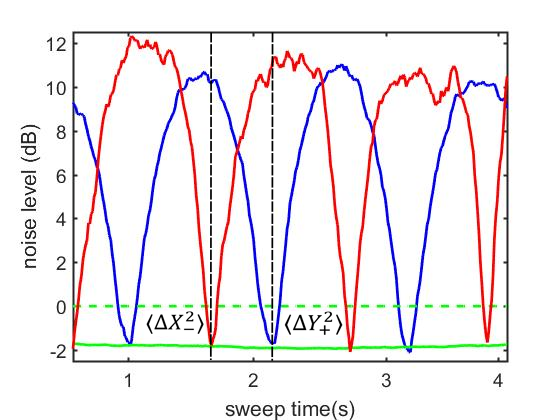}
	\caption{ Joint-quadrature noise as a function of the relative phase $\phi(t)$ as it is varied  from $0$ to $\pi/2$ near the region of best squeezing for each quadrature, $\langle \Delta X_-^2\rangle$ for the red curve and $\langle \Delta Y_+^2\rangle$ for the blue curve. We use the same color scheme as in Fig.~\ref{fig:SU11}(c). The dashed green line shows the shot noise level. The solid green line shows the noise trace when the phase $\phi(t)$ was locked to the value corresponding to the optimal squeezing in the case of interferometric operation ($\phi_{\alpha}=\phi_{\beta}+\pi$, blue curve).}
	\label{fig:quad}
\end{figure}

When the relative phase of the Stokes LO field is offset by $\pi$ with respect to that of the probe field ($\phi_\alpha=\phi_\beta+\phi$), the BPhD performs classical interferometric measurements, as the interference fringes for both probe and Stokes pairs are synchronized.  The probe and Stokes optical fields share maxima at each port. As a result, the differential intensity changes with $\phi(t)$ as $\langle i_- \rangle = 2\alpha \alpha_0 \cos(\phi(t)-\phi_\alpha)$, as shown by blue dashed track in Fig~\ref{fig:SU11}(c). Additionally, the measured noise quadrature moves through the phase-sum and intensity-sum joint-quadratures as the phase $\phi(t)$ sweeps between $0$ and $2\pi$. 

\begin{equation}\label{eq:phase_pi}
\begin{split}
\langle\Delta \hat i^2\rangle =&\alpha^2[\cos^2(\phi(t))\langle(\delta \hat X_{\alpha_0}+\delta \hat X_{\beta_0})^2\rangle\\
&+\sin^2(\phi(t))\langle(\delta \hat Y_{\alpha_0}+\delta \hat Y_{\beta_0})^2\rangle]
\end{split}
\end{equation}

The quadrature-noise curve in Fig.~\ref{fig:SU11}(c) shows the full sweep of the LO phases from 0 to $2\pi$ for both quadrature measurements, and helps visualize how the joint-quadrature quantum noise changes relative to the classical differential intensity. The two joint-quadrature sweeps are out of phase by $\pi/4$, as expected. It is also important to note that when the probe and Stokes intensities at each port are equal, we see the best squeezing. This also happens to be the point where the interference pattern is at its highest sensitivity. This is ideal for phase sensitive measurements.

Fig.~\ref{fig:quad} shows slower and more detailed quadrature noise sweeps, for which the minimum detected noise demonstrates the best squeezing of $-1.8$~dB in both squeezed joint-quadratures. For these measurements, the sweep of the phase ($\phi(t)$) was reduced from $2\pi$ to $\pi/2$, to be able to measure the best squeezing point for each respective quadrature. The sweep had to be reduced due to limitations on phase stability of the system and speed of the spectrum analyzer. The solid green trace was obtained by locking the phase of the LOs relative to the squeezed fields at the position of best squeezing ($\phi_\alpha=\phi_{\alpha_0}+\pi/2$) in the case of $\phi_\alpha=\phi_\beta+\pi$. These phases correspond to the intensity difference signal ($\hat i_-$) being that of a classical interferometer and the sweep being locked to the position of greatest interferometric sensitivity ($\phi_\alpha=\phi_{\alpha_0}+\pi/2$, corresponding to $\langle Y_+^2 \rangle$), shown in Fig.~\ref{fig:SU11}(c) by the minimum and zero point of the solid and dashed blue curves, respectively. As for the shot noise for the system, it was measured by blocking the squeezed rail after the cell allowing for the measurement of the noise powers of the probe and Stokes LOs.

\begin{figure}[htbp]
	\centering
	\includegraphics[width=1\columnwidth]{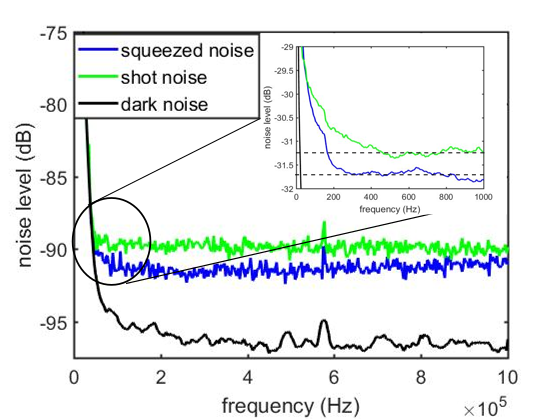}
	\caption{The measured noise spectra for the shot noise (green), squeezed noise (blue), and dark electronic noise (black). The inset zooms shows the low-frequency part of the quantum noise measurements, obtained by taking the Fourier transform of the direct output from the BPhD on an oscilloscope.}
	\label{fig:low_freq}
\end{figure}

One of the clear advantage of the polarization-based realization of the interferometer is its broad-band performance. Since the upper and lower rail probe and Stokes fields mix at the same balanced detector, it makes it easier to balance the two channels and thus reduce the technical noises, which has been a known problem in the original trancated SU(1,1) interferometer ~\cite{LettOE2018}. 
Because of the improved balancing, we are able to achieve uniform joint-quadrature squeezing in broad range of the detection frequencies and naturally extend it to low-frequency detection, as shown in Fig.~\ref{fig:low_freq}. The main figure shows a nearly flat sub-shot noise squeezing level for the $X_-$ joint quadrature from 1~MHz down to 50~kHz (limited by the $10$~kHz) spectrum analyzer resolution bandwidth). For more accurate squeezing measurements at the low-frequency region, we have directly recorded the BPhD using an oscilloscope, and the inset in Fig.~\ref{fig:low_freq} shows the Fourier transform of this signal. This detection method allowed squeezing observation in sub-kHz detection frequencies ($\le 200$~Hz), below which the signal was contaminated by power line noise. With better electronic noise isolation we may be able to reduce the detection bandwidth even further, which makes this system particularly attractive for direct quantum noise imaging. 

We can also use these joint-quadrature noise measurements to characterize the degree of quantum entanglement between the probe and Stokes fields. One way to characterize the entanglement is the inseparability criteria~\cite{lettSci08,PhysRevLett.84.2722}: \\
%
$I=\langle\Delta X_-\rangle^2+{\langle\Delta Y_+\rangle}^2\le2$.
%
For the joint-quadrature measurements, shown in Figs.~\ref{fig:SU11}(c) and ~\ref{fig:quad}, the best squeezing values are $\langle\Delta X_-\rangle^2=\langle\Delta Y_+\rangle^2=.66\pm.03$, giving an inseparability parameter value of $I=1.32\pm.04<2$. This allows us to claim that the twin beams are entangled. However, these results are not sufficient to satisfy the more strict EPR entanglement criteria:\\
$4\langle\Delta X_-\rangle^2\langle\Delta Y_+\rangle^2\le1$,
\\
and the minimum value of this parameter we were able to achieve in our experiment was $\sim$1.75 for the measured conditional variance.

It is important to note that the detected squeezing values were severely affected by the optical losses after the cell (~30\%), some thermal jitter, the differential local oscillator phase, and leakage of the pump field into the measured signals. All these problems are technical and can be resolved with higher-quality optical elements. 
For example, if we just account for the optical losses using the beam splitter model, the inferred level of squeezing would be $-2.9$~dB, corresponding to inseparability value of $I\simeq 1$. Moreover, we were able to measure up to $-4.5$~dB of two-mode intensity squeezing in a single rail, that lets us assume that with better phase stabilization and pump filtering we should be able to achieve similar levels of the joint-quadrature squeezing in the polarization-based SU(1,1).

It is also important to note that such a polarization-based truncated SU(1,1) interferometer can be used to generate polarization entangled states~\cite{bell_states}. By removing the seed probe fields completely, and using vacuum input for both rails, we should be able to generate full set of the polarization-entangled Bell states which will have applications in quantum communications and can be applied to other phase sensitive measurement techniques~\cite{Chekhova:16AOPreview}.

In conclusion, we have demonstrated a modification of the truncated SU(1,1) interferometer with full optical control over the output joint quadratures. Our design utilized four-wave mixing nonlinearity in thermal Rb vapor, but can be streightforwardly adopted for a broad range of nonlinear systems~\cite{int_reviewPRA1986,Chekhova:16AOPreview}. In our experiments we achieved up to $-1.8$~dB of joint-quadrature twin beam squeezing, proving their quantum inseparability. The squeezing value can be further improved by reducing optical losses, improving pump field filtering, and adding active stabilization of the interferometer. The demonstrated interferometer design provides intrinsically more symmetric detection, resulting in better balancing and consequently better technical noise suppressions, especially at lower detection frequencies. In the current setup we were able to observe joint-quadrature squeezing at frequencies as low as 200 Hz to up to a few MHz. Thus, the proposed setup has many potential applications in quantum imaging and quantum metrology. Moreover, it can be immediately applied for generation of polarization-entanlged Bell states by seeding both probe inputs with vacuum. 

National Science Foundation (NSF) (Phy-308281).  

We would like to thank Kelly Roman, Haley Bauser and Nathan Super for their contributions to the experimental setup developments, and Alberto Marino, Eugeniy Mikhailov, and Meng-Chang Wu for useful comments and suggestions.

\end{document}